\documentclass[%
 reprint,
superscriptaddress,
 amsmath,amssymb,
 aps,nofootinbib,
]{revtex4-1}

\usepackage{graphicx}
\usepackage{dcolumn}
\usepackage{aasjournal} 
\usepackage{bm}
\usepackage{xcolor}

\begin{document}


\title{The effective volume of supernovae samples and sample variance}

\author{Zhongxu Zhai}
  \email{zhongxuzhai@sjtu.edu.cn}
\affiliation{Department of Astronomy, School of Physics and Astronomy, Shanghai Jiao Tong University, Shanghai 200240, China}
\affiliation{Shanghai Key Laboratory for Particle Physics and Cosmology, Shanghai 200240, China}
\affiliation{%
Waterloo Center for Astrophysics, University of Waterloo, Waterloo, ON N2L 3G1, Canada}
\affiliation{
Department of Physics and Astronomy, University of Waterloo, Waterloo, ON N2L 3G1, Canada
}%

\author{Will J. Percival}%
  \altaffiliation[Also at ]{Perimeter Institute for Theoretical Physics, 31 Caroline St. North, Waterloo, ON N2L 2Y5, Canada}

\affiliation{%
Waterloo Center for Astrophysics, University of Waterloo, Waterloo, ON N2L 3G1, Canada}
\affiliation{
Department of Physics and Astronomy, University of Waterloo, Waterloo, ON N2L 3G1, Canada
}%

\author{Zhejie Ding}
\affiliation{Department of Astronomy, School of Physics and Astronomy, Shanghai Jiao Tong University, Shanghai 200240, China}
\affiliation{Shanghai Key Laboratory for Particle Physics and Cosmology, Shanghai 200240, China}

\date{\today}

\begin{abstract}
The source of the tension between local SN Ia based Hubble constant measurements and those from the CMB or BAO+BBN measurements is one of the most interesting unknowns of modern cosmology. Sample variance forms a key component of the error on the local measurements, and will dominate the error budget in the future as more SNe Ia are observed. Many methods have been proposed to estimate sample variance in many contexts, and we compared results from a number of approximate methods to estimates from N-body simulations in a previous paper, confirming that sample variance for the Pantheon SNe Ia sample does not solve the Hubble tension. We now extend this analysis to include the more accurate analytic method based on calculating correlations between the radial peculiar velocities of SNe Ia, comparing this technique with results from numerical simulations. 
We consider the dependence of these errors on the linear power spectrum and how non-linear velocities contribute to the error. Using this technique, and matching sample variance errors from more approximate methods, we can define an effective volume for SNe Ia samples, finding that the Pantheon sample is equivalent to a top-hat sphere of radius $\sim220~h^{-1}$Mpc. We use this link between sample-variance errors to compute $\Delta H_{0}$ for idealised surveys with particular angular distributions of SNe Ia. For example, a half-sky survey at the Pantheon depth has the potential to suppress the sample variance of $H_{0}$ to $\sim0.1$ km s$^{-1}$Mpc$^{-1}$, a significant improvement compared with the current result. Finally, we consider the strength of large-scale velocity power spectrum required to explain the Hubble tension using sample variance, finding it requires an extreme model well beyond that allowed by other observations.

\end{abstract}

\maketitle


\section{Introduction} 

The determination of the Hubble constant using the local distance ladder relies on a sample of Type Ia Supernovae (SNe Ia) with high quality \cite{localH0-2016,localH0-2021}. Since each SN Ia observation probes the space-time along the line of sight (LoS), inhomogeneities between the observer and the SN Ia alter the value of $H_{0}$ recovered. For a set of SNe Ia, these distortions introduce sample variance in the final measurement of $H_{0}$. This effect can be reduced when we increase the number of SN Ia observations in different directions. Instead of considering individual LoS, we can take a holistic view considering that the SNe Ia cover a patch in the Universe, and that this patch does not follow the behaviour of the background. By considering density fluctuations on scales larger than the patch, we can calculate the expected variance of parameters such as $H_{0}$ between a set of patches. These methods - looking at perturbations along each line-of-sight or looking at perturbations in patches of the universe - were compared in an earlier paper \cite{Zhai_2022} and shown to give consistent results. When the current compilation of SNe Ia is taken into account, we found that the sample variance error is around $\sim0.4$ km s$^{-1}$Mpc$^{-1}$, not able to explain the tension of $H_{0}$ using local distance ladder and Cosmic Microwave Background, as has been found previously \cite{Marra2013,Wojtak2014,Romano2016,Wu2017,Camarena_2018,Dainotti_2021,Dainotti_2022}.

Although the amplitude of sample variance is not sufficient to explain the Hubble tension, it still contributes a significant component of the error budget of the $H_{0}$ determination, given the fact that the latest measurement has reached a combined uncertainty level of 1 km s$^{-1}$Mpc$^{-1}$. All methods for calculating sample variance start from the same step, integrating the linear power spectrum to quantify the amplitude of density fluctuations. The SNe Ia living in an overdense area experience an additional attraction due to local structures and reduce the $H_{0}$ estimate, while the opposite can happen in underdense regions and increase $H_{0}$. As discussed above, the behaviour of these regions can be considered as either giving rise to peculiar velocities of SNe Ia with respect to the background, or changing the cosmology of the overdense/underdense patches such that there are no peculiar velocities within the patch, just each patch of space-time is behaving in a different way. 

The first method considered in \cite{Zhai_2022} (hereafter paper I) was based on how the inhomogeneity $\delta$ along the LoS changes the luminosity distance in a frame where the redshift remains constant  \cite{Sasaki1987,Barausse2005,Bonvin2006,Hui_2006}. Relativistic corrections to observations in such a model were considered in  \cite{Fonseca_2023}. In the local universe, the resulting variance in $H_{0}$ can be approximated as $-f\delta/3$ where $f$ is the linear growth rate. The second method considered in paper I was based on methods to determine super sample covariance (SSC) in sets of cosmological simulations \cite{Frenk88,Sirko2005,Baldauf_2011}. Due to the finite volume covered in any simulation, density fluctuations on scales larger than the simulation box give rise to a ``DC-level" density fluctuations that is different for each box in a set. This then leads to cosmological parameters that vary between the simulations \cite{Sirko2005}. The Hubble constant in each box (or patch) is different from the background and measurements within it will give this local value. The third method borrows ideas from the homogeneous top-hat model for structure growth. In a small patch of the universe, the evolution can be governed by the same equations as the background but with a different initial curvature due to perturbations in density. Each sphere is governed by a Friedmann equation with different parameters, leading to a different scale factor inside the patch than the background and thus a different $H_{0}$. The fourth method uses numerical simulations directly - in N-body simulations, distances are measured relative to the background, with perturbations resulting in the peculiar velocities that incorporate the dynamics of dark matter. 

The three analytical methods described above perturb different parameters in the model: the luminosity distance, cosmological parameters or spatial curvature, with the level of perturbation limited by the density power spectrum. The simulation-based method doesn't perturb any parameters in the model explicitly, but the correlation of $H_{0}$ with the overdensity in the patch $\delta$ can be easily computed from the peculiar velocities within the simulation. In this paper, we contrast these methods with a method based on directly modeling the correlations between the radial peculiar velocities of different SNe Ia. In a frame where distances are measured with respect to the background cosmology, and perturbations affect peculiar velocities, the sample variance of $H_{0}$ results from the correlated peculiar velocities of the SNe Ia. Such a method was introduced in earlier works \cite{Gorski_1988, Gorski_1989} and has been studied in the constraint on cosmological parameters, see \cite{Jaffe_1995, Borgani_2000, Hui_2006, Nusser_2011, Davis_2011, Hudson_2012, Okumura_2014, Howlett_2017, Wang_2018, Boruah_2020} and references therein. Similar to the non-simulation methods summarized above and in paper I, the velocity correlation function method can also be classified as an analytical method. However, there are crucial differences. First, this method utilizes the peculiar velocity of the SN Ia directly, which is the source of the sample variance in the $H_{0}$ measurement. For methods that perturb parameters, one has to approximate the size and shape of the patch being considered. Instead, by considering the peculiar velocity field directly as defined within a single background cosmology, this method uses the exact geometry of the SN Ia survey. Thus, it provides more accuracy. When originally defined, the method was used to predict the peculiar velocities of dark matter halos leading to results for peculiar velocity surveys and Redshift Space Distortion (RSD) measurements, matching simulation-based analysis. The method gives the effect of sample variance (or RSD) on SN Ia surveys, but the link from peculiar velocity errors to errors on $H_0$ is not direct - the method to go from the velocity errors to errors on the measurement of H0 requires care, as discussed later, where we use Monte-Carlo simulations to determine this link.

Each of the five methods used to estimate sample variance for SNe Ia measurements provides different insights for the method. By comparing results between direct measurements that include the 3D distribution of SNe Ia and those that approximate this as a simple shape, we can define effective properties of any sample. This effective volume makes it possible to compare and contrast surveys. Ideally, we want to use fast analytic methods to make predictions, but need to do so accurately. The problem with using the analytic methods in paper I directly is that they make too many simplifying assumptions to be accurate. For example, that each SN Ia in the sample has equal weight, and depending on the exact method, they make an approximation such as that the region of influence of every line-of-sight (LOS) to a SN Ia is a sphere entered on the mid-point on the LOS between SN and the observer. However, the SNe Ia at higher redshift have a larger contribution from the Hubble flow than their peculiar velocity, and therefore in terms of the $H_{0}$ determination, they should receive more weight. In addition, the sample variance of $H_{0}$ is directly affected by the uncertainty of the apparent magnitude of each SN Ia, see e.g. \cite{Wu2017} and thus it's not correct to assume that they are the same given the noise covariance matrix \cite{Pantheon}.

We therefore need to consider a more realistic method to estimate the effective volume of the SNe sample. By matching the variance $\Delta H_{0}$ of the approximate methods to a more accurate method, we can achieve this goal and define an effective volume for any sample of SNe Ia in a fast, quantitative, and straightforward way. We could do this calibration using N-body simulations, but this is computationally very expensive. Instead we now show that we can calibrate the fast methods of paper I using the velocity correlation function method. Based on this, we can quickly forecast the sample variance component of $\Delta H_{0}$ for future SNe Ia surveys as a function of area and number density, and balance the breadth of any survey against density for sample variance errors.

This paper is organized as follows, Section~II presents the formalism of the radial velocity correlation function. Section-III provides the results for the estimate of $\Delta H_{0}$ and the effective volume of the current SNe data. Section~IV includes our discussion and conclusions. 

\section{Velocity correlation function} \label{sec:vcf}

In this section, we briefly review the formalism of the velocity correlation function method developed in the literature, and describe how we apply it to estimate the $H_{0}$ variance.

For a set of SNe Ia, the sample variance depends on the relative position of the LoS through the radial velocity correlations e.g. \cite{Gorski_1988, Davis_2011, Hui_2006, Wang_2018, Wang_2021}. For example, probing similar LoS multiple times does not reduce the sample variance as much as probing widely separated LoS. In a frame where perturbations manifest as peculiar velocities, the radial velocity correlation function shows how correlated the sample variance errors are between any two SNe Ia. 

For multiple SNe Ia we can construct a $n\times n$ matrix, where $n$ is the number of SNe Ia, and each element is the covariance between the radial velocities of two SNe Ia. Given such a matrix, and assuming that the velocities are drawn from a Gaussian distribution, we can easily create Monte-Carlo samples of sets of peculiar velocities. These can then be attached to measured SNe Ia positions and $H_0$ can be measured for each realisation, using the same methodology applied to data. The distribution of $H_0$ recovered then leads to the sample variance error on $H_0$. This method can be considered an approximation to N-body simulations, with the simulations evolving the field in order to get the correct non-linear velocities for the distribution of SNe Ia positions, while the matrix uses a known form for the covariance within a Gaussian random field. 

The method starts from the velocity correlation function, which is defined as 
\begin{equation}
\Psi_{i,j}(\bf{r}) \equiv \langle v_{i}(\bf{r_{a}})v_{j}(\bf{r_{b}})\rangle,
\end{equation}
where $\mathbf{r}=\mathbf{r_{b}}-\mathbf{r_{a}}$ is the separation vector of two host galaxies labeled as ``a" and ``b",  $i$ and $j$ denote the Cartesian components of the velocity and the average is over all the galaxy pairs. In practice, the SN Ia measurements only depend on the radial component of the velocity, and it is convenient to define ${\bf u_{a}}=\hat{\bf r}_{a}u_{a}=\hat{\bf r}_{a}(\hat{\bf r}_{a}\cdot \bf{v_{a}})$ and ${\bf u_{b}}=\hat{\bf r}_{b}u_{b}=\hat{\bf r}_{b}(\hat{\bf r}_{b}\cdot {\bf v_{b}})$, where $\hat{\bf{r}}$ is the unit direction vector for $\bf{r}$. 

For a statistically isotropic and homogeneous random vector field, \cite{Gorski_1988} showed that the above velocity correlation tensor can be written as a function of the amplitude of the separation vector $r=|\bf{r}|$
\begin{equation}
    \Psi_{i,j}(r) = \Psi_{\parallel}(r)\hat{r}_{i}\hat{r}_{j}+\Psi_{\perp}(r)(\delta_{ij}-\hat{r}_{i}\hat{r}_{j}),
\end{equation}
the two new functions $\Psi_{\parallel}$, $\Psi_{\perp}$ are the radial and transverse components of the velocity correlation function, respectively. For the radial peculiar velocities, we can write the correlation function as 
\begin{eqnarray}
 \langle u_{a}u_{b}\rangle &=& \hat{r}_{ai}\hat{r}_{bj}\langle v_{i}v_{j}\rangle = \Psi_{\parallel}(r)(\hat{\bf{r}}_{a}\cdot\hat{\bf{r}})(\hat{\bf{r}}_{b}\cdot\hat{\bf{r}}) \\
    &&+\Psi_{\perp}(r)[\hat{\bf{r}}_{a}\cdot\hat{\bf{r}}_{b}-(\hat{\bf{r}}_{a}\cdot\hat{\bf{r}})(\hat{\bf{r}}_{b}\cdot\hat{\bf{r}})].
\end{eqnarray}
This expression can be further simplified in terms of the angles $\theta_{1}$ and $\theta_{2}$ between the separation vector $\bf{r}$ and the galaxy position vectors $\bf{r_{a}}$ and $\bf{r_{b}}$, $\cos\theta_{a}=\hat{\bf{r}}_{a}\cdot\hat{\bf{r}}$, $\cos\theta_{b}=\hat{\bf{r}}_{b}\cdot\hat{\bf{r}}$, and $[\hat{\bf{r}}_{a}\cdot\hat{\bf{r}}_{b}-(\hat{\bf{r}}_{a}\cdot\hat{\bf{r}})(\hat{\bf{r}}_{b}\cdot\hat{\bf{r}})]=\sin\theta_{a}\sin\theta_{b}$. In this case, the velocity correlation function is fully determined by the functions $\Psi_{\parallel}$ and $\Psi_{\perp}$. In linear theory, Ref. \cite{Borgani_2000} shows that they can be computed through the power spectrum of the density fluctuation $P(k)$ as follows:
\begin{eqnarray}  \label{eq:Psi_par}
    &&\Psi_{\parallel}=\frac{(fH_{0})^{2}}{2\pi^{2}}\int P(k)\left[j_{0}(kr)-2\frac{j_{1}(kr)}{kr}\right]dk, \\
    &&\Psi_{\perp} = \frac{(fH_{0})^{2}}{2\pi^{2}}\int P(k)\frac{j_{1}(kr)}{kr}dk, \label{eq:Psi_perp}
\end{eqnarray}
where $f$ is the logarithmic derivative of the linear growth factor with respect to the scale factor and can be approximated as $f\approx\Omega_{m}^{\gamma}$ with $\gamma$ as the growth index \cite{Wang_1998, Lue04}, and $j_{n}$ is the spherical Bessel functions of order $n$. The power spectrum $P(k)$ can be easily computed using models like the parameterization of \cite{Eisenstein_1998}. The above equations can determine the velocity correlation function of galaxy pairs. In the limit where $r\to0$, Eqns.~\ref{eq:Psi_par} \&~\ref{eq:Psi_perp} reduce to the same form, which gives the diagonal elements in the covariance matrix, 
\begin{equation}\label{eq:dispersion}
    \sigma^2_{v_{i}} = \langle u^{2}\rangle = \frac{(fH_{0})^{2}}{6\pi^{2}}\int P(k)dk,
\end{equation}
where $\sigma_{v_{i}}$ is the dispersion in the peculiar velocities \cite{Hui_2006}. Thus we have completed the construction of the velocity correlation function for the SNe Ia sample. Note that Eq.~\ref{eq:Psi_par}~\ref{eq:Psi_perp} \&~\ref{eq:dispersion} are only valid for linear scales, where the density-velocity relationship is simple. To extend this into the non-linear regime, we would need to use the non-linear velocity power spectrum, rather than scaling the non-linear density power. For instance, \cite{Howlett_2019} explores the theoretical descriptions for modelling the non-linear effect on the velocity correlation function in redshift space. Although our linear model is clearly too simple, the results given below show good agreement with the simulation-based analysis. The velocity correlation function method has been previously used to predict errors for a SN Ia sample. For instance, \cite{Huterer_2015} adopt a likelihood analysis to search the significance of the bulk velocity using the JLA sample at low redshift. And \cite{Castro_2016} uses the same SNe Ia sample to retrieve constraint on the growth-related cosmological parameters using SN peculiar velocity with and without combining the weak lensing signal. We extend this work, comparing the correlation function method to approximate methods and simulations, and using the comparison to define the effective volume of a SN sample. Using this effective volume we can easily make predictions for future projects that would otherwise be computationally prohibitive.

\section{Results}

In this section, we use the velocity correlation function method to estimate the sample variance of $H_{0}$ from the current SNe Ia dataset, and investigate extensions including non-linear corrections to the power spectrum and velocity bias.

\subsection{Estimating the Variance of $H_{0}$}

\begin{figure}
\begin{center}
\includegraphics[width=8.5cm]{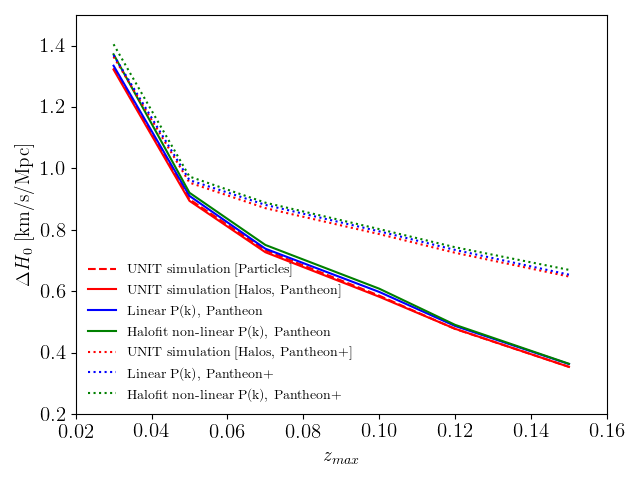}
\caption{Variance of $H_{0}$ measurements estimated using purely simulation-based method (solid red) and new method based on velocity correlation function (solid blue). For comparison, the dashed red line shows result when the SNe Ia are matched to dark matter particles instead of halos, and the solid green line represents result when the power spectrum for the calculation of velocity correlation function considers non-linear correction to the density power using halofit, coupled with a linear relationship between density and velocity power spectra.}
\label{fig:DeltaH0_VCF}
\end{center}
\end{figure}

We compare results from the velocity correlation method against those from numerical simulations calculated using method~D in paper I. Our numerical simulation based estimate of the $H_{0}$ sample variance relies on a large-scale N-body simulation from UNIT \footnote{\url{http://www.unitsims.org/}} \cite{Chuang2019}. We use the halo catalog from the simulation, and randomly choose a halo with a mass of $\sim10^{12-15}h^{-1}M_{\odot}$ as the position of the observer. Then we consider the Pantheon compilation \cite{Pantheon} of the SNe Ia sample within the redshift range $0.023<z<z_{max}$ where $z_{max}$ is the maximum redshift in the local distance ladder measurement and the fiducial analysis in \cite{localH0-2021} adopts $z_{max}=0.15$. We assign each SN to the nearest dark matter halo and in this case, the peculiar velocity of the dark matter halo is inherited by the SNe Ia. We can measure $H_{0}$ through
\begin{equation}
    \log{H_{0}} = 0.2M_{B}^{0}+a_{B}+5,
\end{equation}
where $M_{B}^{0}$ is the fiducial luminosity of SNe Ia and $a_{B}$ is the expansion parameter describing luminosity distance and redshift relation. The $H_{0}$ uncertainty contributed from SNe Ia peculiar velocity is via the uncertainty of 
\begin{equation}\label{eq:Delta_ab}
    \Delta a_{B} = \frac{1}{N}\sum_{i=1}^{N}\frac{1}{\ln{10}}\frac{v_{i}}{r_{i}H_{0}},
\end{equation}
where $v_{i}$ is the the peculiar velocity in the radial direction of the $i-$th SN Ia, and $N$ is the total number of SNe Ia used in the analysis \cite{Wu2017}. We estimate the final variance $\Delta H_{0}$ by repeating the process $10^{4}$ times to get a distribution, i.e. each iteration has a different observer in the simulation box and we randomly rotate the SNe Ia sample as a whole into different directions.

For our new method based on the velocity correlation matrix, we can use the observed SN Ia positions as the starting point to estimate the velocity covariance matrix. However, in order to utilise the existing routines for analysing simulations, and to compare more closely to the results from N-body simulations, we modify this slightly and apply the same procedure to assign SNe Ia to dark matter halos and then we construct a covariance matrix for peculiar velocities using the formula in Section \ref{sec:vcf} integrating over the expected power spectrum \footnote{If, instead, we had used the angular position and redshift of the SNe Ia sample to construct the velocity correlation matrix, this would be kept fixed for all realisations. When we match SNe Ia to the halo catalog, it causes slight variations of the positions in each iteration. Our test shows that the difference of these two methods is less than 5\%. We fit to the simulations for both in order to have a cleaner match between the two methods}. The covariance matrix only depends on the relative positions of the SNe Ia (or halos). We then use this covariance matrix to produce a mock data vector for peculiar velocities sampling from a multi-variate Gaussian distribution, and assign them to the halos. We repeat this process $10^{4}$ times to estimate $\Delta H_{0}$.

In Fig.~\ref{fig:DeltaH0_VCF}, we present $\Delta H_{0}$ as a function of $z_{max}$. We can see that our new method of modelling peculiar velocity (solid blue) produces results in excellent agreement with our previous purely simulation-based method (solid red). The relative difference between the two method is no higher than a few percent. They both show a clear monotonic dependence on $z_{max}$, i.e. higher $z_{max}$ means more and distant SNe Ia are added in the analysis of distance ladder. For the current fiducial analysis with $z_{max}=0.15$, the sample variance of $H_{0}$ measurement is lower than $\sim0.4~\text{km}~\text{s}^{-1}\text{Mpc}^{-1}$. The new method further approves the robustness of this estimate and shows that the sample variance itself is not able to fully resolve the tension with Planck.

\subsection{Nonlinear correction}

Our model of the velocity correlation function is built upon the linear perturbation theory. A fully non-linear description requires a model for the non-linear velocity power spectrum, which may in turn require a detailed analysis using high resolution simulations. However, it is possible to perform a simple approximate analysis using the non-linear density power spectrum and assuming a linear relationship between density and velocity (such a relationship is often assumed in an analysis of cosmic voids \cite{Woodfinden_2022}). To do this, we apply the halofit model \cite{Takahashi_2012} for non-linear power spectrum $P(k)$ and redo the estimate of $\Delta H_{0}$. The result is shown as solid green line in Fig.~\ref{fig:DeltaH0_VCF}. Since the non-linear correction can boost the power spectrum at small scales, the immediate impact is to increase the velocity dispersion (Eq. \ref{eq:dispersion}), i.e. the diagonal elements for the covariance matrix, which increases the estimate $\Delta H_{0}$. However, the comparison with the linear model shows that the impact is quite small and the overall agreement with other methods remains. In addition, one can use models like \cite{Howlett_2019} to describe the non-linear correction in a more realistic manner, however the agreement between our approach and the simulation-based method seems to show that this effect is negligible for the resulting $H_{0}$ variance.

\subsection{Velocity bias}

As a byproduct of our analysis for peculiar velocity, we can investigate the large-scale velocity bias - the link between the halo and matter velocity fields - using numerical simulations. Earlier studies have shown that this parameter is close to unity, i.e. the galaxy/halo has a velocity field close to the underlying dark matter field \cite{Desjacques_2010,Chen_2018, Zhang_2018}. These analyses compare the two-point statistics for galaxies/halos with dark matter particles and explored any dependence of velocity bias on halo mass or redshift. 

Our estimate of sample variance for $H_{0}$ using numerical simulations can be easily extended to study velocity bias. We replace halos by dark matter particles in the assignment of SNe Ia to halos. In Fig.~\ref{fig:DeltaH0_VCF}, the dashed red line displays this result. It shows that the two velocity field are exactly the same in terms of $\Delta H_{0}$, indicating that the velocity bias of dark matter halos is close to unity, similar to the results from literature. The work in \cite{Chen_2018} shows some deviation of velocity bias from unity at $z=0$ for small scale, but the deviation is no higher than 5\%. Therefore our result is not in significant conflict with theirs, but can serve as an independent measurement of velocity bias. 

\subsection{Re-analysis with the Pantheon+ compilation}

\begin{figure}
\begin{center}
\includegraphics[width=8.5cm]{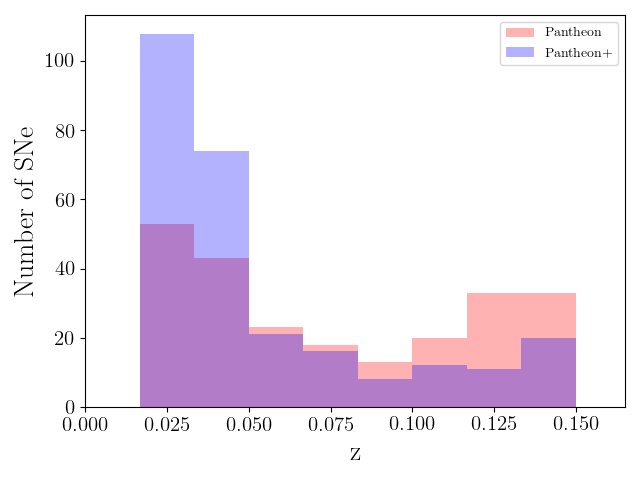}
\caption{The redshift distribution of Pantheon and Pantheon+ SNe Ia samples that have been used to measure $H_{0}$, in the redshift range $0.023<z<0.15$}
\label{fig:pantheon+_redshift}
\end{center}
\end{figure}

The results discussed above were based on the Pantheon data, and we note that this catalog was recently updated to Pantheon+, as described in \cite{Scolnic_2022}, and used to remeasure $H_{0}$ in \cite{localH0-2021}. We now compare the two samples and the sample variance in the new $H_{0}$ measurement for Pantheon+. We estimate the sample variance error using the velocity correlation function method, and present the results in Figure \ref{fig:DeltaH0_VCF}, shown as the dotted lines. The first feature is the agreement between the velocity correlation function method and the simulation based method, which matches the agreement found using the Pantheon data. This provides further evidence of the method used in this work. However, there is substantial offset between Pantheon and Pantheon+ at higher $z_{max}$. This is counter-intuitive since Pantheon+ has more SNe Ia than Pantheon. The reason is that Pantheon+ is not a simple extension to Pantheon. The selections and cuts in the new compilation rearrange the catalog significantly, see e.g. the discussions in \cite{Scolnic_2022} and \cite{Carr_2022}. In the redshift range $0.023<z<0.15$ for the $H_{0}$ measurement, the overlap between Pantheon and Pantheon+ is less than 50\%. More importantly, Pantheon+ has more SNe Ia at low redshift, but fewer at high redshift as shown in Figure \ref{fig:pantheon+_redshift}. The Hubble flow is more significant for individual SN Ia compared with the contribution from peculiar velocity, therefore the high redshift SNe Ia gain more weight in the $H_{0}$ determination. At $z>0.1$, the results from two compilations differ by a factor of two. Although the Pantheon+ has more data, the sample variance in the $H_{0}$ measurement is not suppressed. From this point of view, the selection is not optimized, implying that future improvement of the variance could be possible.

\subsection{Effective volume of SNe Ia sample}

\begin{figure}
\begin{center}
\includegraphics[width=8.5cm]{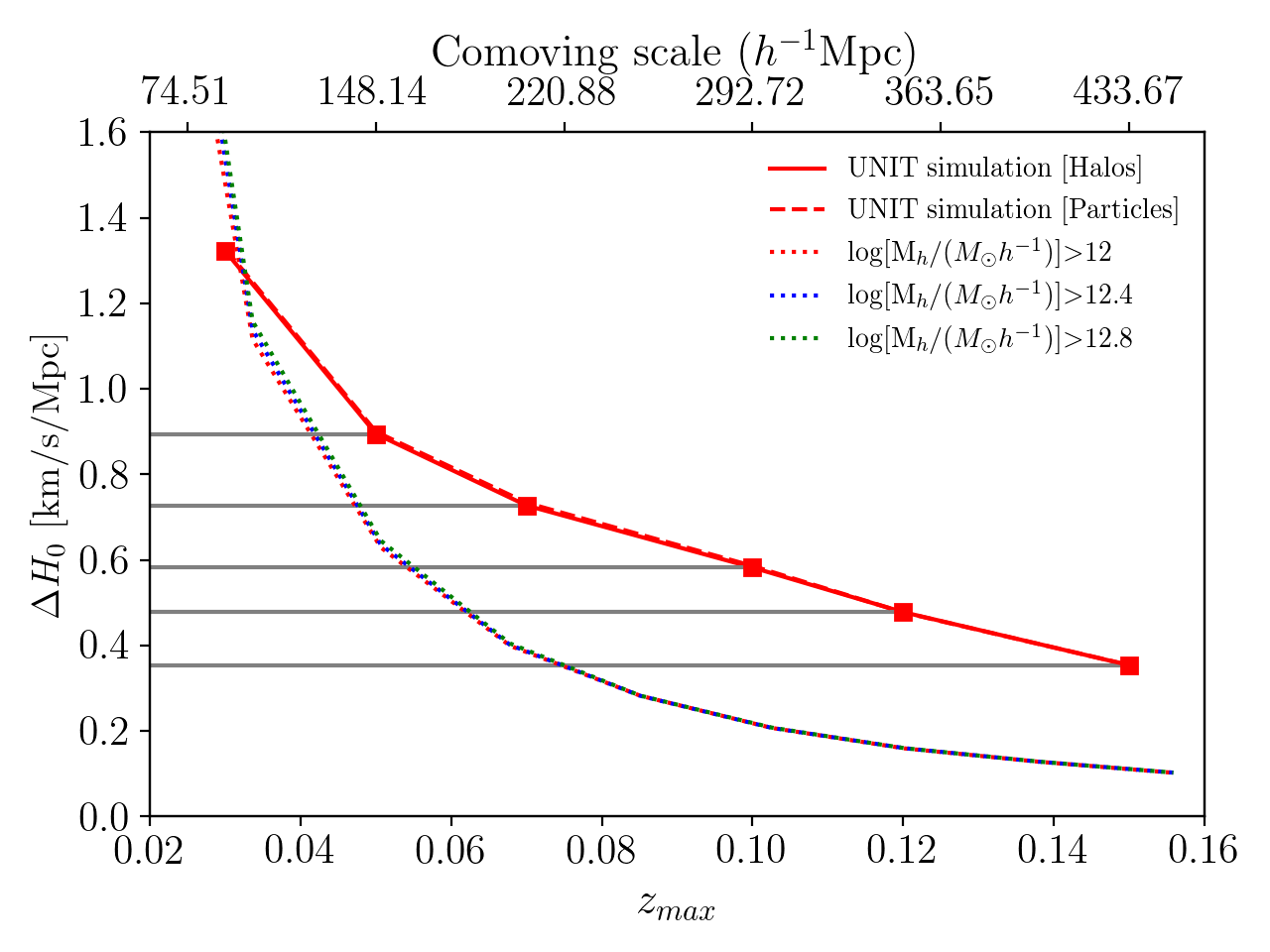}
\caption{Estimate of the effective volume of the SNe Ia sample. The solid and dashed red lines represent $\Delta H_{0}$ when we match SNe Ia to dark matter halos or particles in the UNIT simulation. The dotted lines denote results when we use all halos above some mass cut within some radius to estimate $\Delta H_{0}$. The interception between the horizontal grey lines and the dotted lines give the effective scale of the SNe Ia sample up to $z_{max}$.}
\label{fig:effective_volume}
\end{center}
\end{figure}

In paper I, we applied a region of influence method and estimated that the SNe Ia sample for the distance ladder measurement only probes a much smaller volume than the maximum redshift $z_{max}=0.15$. By comparing techniques for estimating sample variance errors, we can now adopt a straightforward calculation to find the effective region of influence for a SN Ia sample: to do this, we can use all the halos within radius $R$ and estimate $\Delta H_{0}$ as a function of $R$. Then we find the scale where the result is the same $\Delta H_{0}$ as that calculated by the velocity correlation function method. The resultant scale can be considered as the radius of the effective volume that the SNe Ia sample probes. 

In Fig.~\ref{fig:effective_volume}, we present the result for the Pantheon sample of SNe Ia. The solid and dashed red lines are the same as Fig.~\ref{fig:DeltaH0_VCF}, i.e. $\Delta H_{0}$ from the current Pantheon sample using dark matter halos or particles respectively. The dotted lines stand for the results when we use all halos within some radius. For comparison, we present results with different lower limits of the halo mass. We can see that the result has no dependence on the mass range, matching our previous finding about the velocity bias of dark matter halos. We search the effective volume of the SNe Ia data by matching $\Delta H_{0}$ of the two measurements, i.e. the horizontal grey lines. For $z_{max}=0.15$, we find that the effective volume corresponds to a scale of about $220 h^{-1}$Mpc. This is larger than our earlier estimate in \cite{Zhai_2022}, meaning that in terms of $\Delta H_{0}$, the current distance ladder measurement is equivalent to a set of SNe Ia with the same number, within a spherical patch out to redshift $\sim0.075$. 
If we look at the Pantheon+ sample, we estimate that the effective volume matches that of a spherical patch out to redshift $\sim0.05$, smaller than the Pantheon sample since the data is more concentrated at low redshift.

\subsection{Implications for future SNe Ia survey: ideal case}

\begin{figure*}
\begin{center}
\includegraphics[width=17.5cm]{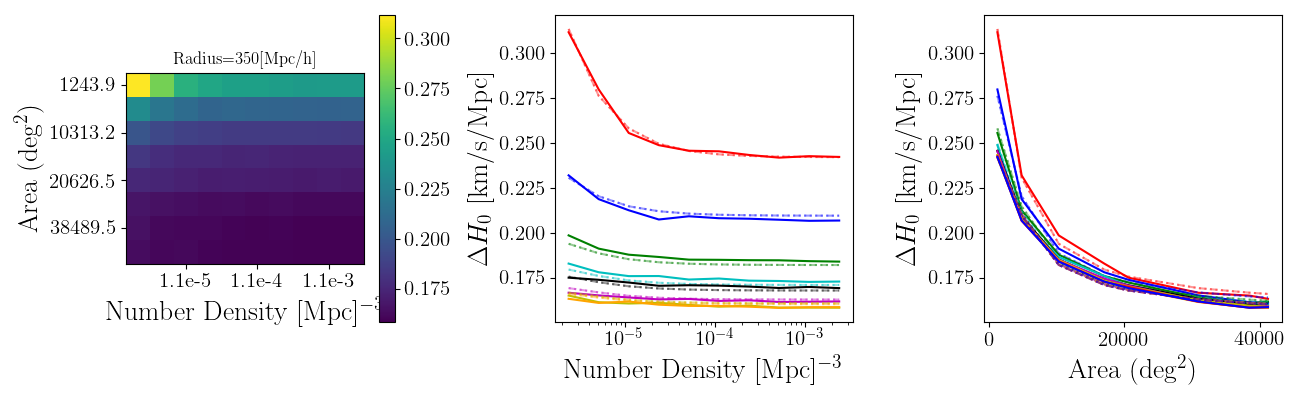} \\
\includegraphics[width=17.5cm]{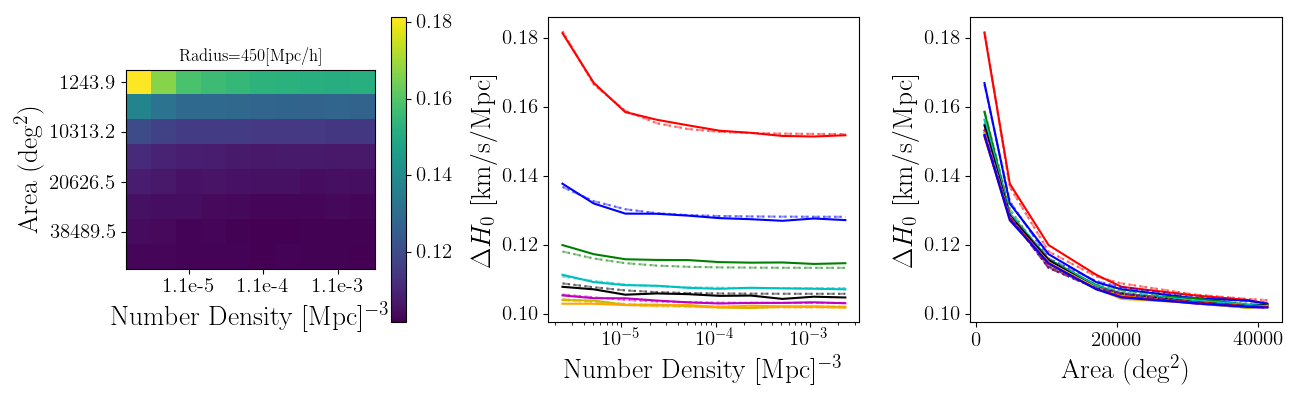}
\caption{Dependence of $\Delta H_{0}$ on the parameters $A$ and $n_{sn}$ assuming an isotropic SNe Ia survey on the sky with an angular coverage of a spherical cap. The left hand panel shows the dependence on the 2D plane, while the middle and right hand panel show the dependence on individual parameters. The different colors in the middle (right) panel shows different values of $n_{sn}$ ($A$). The dotted lines are the best-fit result with a second polynomial fit (Eq. \ref{eq:fitting}).}
\label{fig:for_survey}
\end{center}
\end{figure*}

Although the current error contribution from SNe Ia in the measurement of $H_{0}$ is around $\sim0.4~\text{km}~\text{s}^{-1}\text{Mpc}^{-1}$, subdominant in the total error budget, the accuracy will soon be improved with the LSST and Roman surveys \cite{Dodelson_2016, Villar_2018, Rose_2021}. Using numerical simulations, we are able to explore the impact of survey designs on the sample variance of the $H_{0}$ measurement.

Without any prior knowledge about the survey shape and coverage, we simply assume that the survey is a spherical cap on the sky and the halos are homogeneously distributed within this volume. With this approximation, there are three parameters to define the sample: the survey area $A$, radius $R$ (sometimes defined using $z_{max}$), and halo number density $n_{sn}$. Since the analysis is based on the UNIT simulation, the halo catalog implies an upper limit on the number of objects we can use. For reference, this value is $2.38\times10^{-3}$ [\text{Mpc}]$^{-3}$ for dark matter halos within a mass range of $10^{12\sim15}~h^{-1}M_{\odot}$. We explore the range of $n_{sn}$ that can be as small as 0.1\% of this reference number density. For a given $R$, we define a 2D grid for $A$ and $n_{sn}$, then we compute $\Delta H_{0}$ for each $A$ and $n_{sn}$ using halos that can be selected.


As an example, in Fig.~\ref{fig:for_survey} we present results for $R=350$ and $450 h^{-1}$Mpc. The left panel shows the distribution for a range of $A$ and $n_{sn}$, while the middle and right panels show the dependence on individual parameters. We can see that the overall shape of the curves is similar for different values of $R$, as well as the change as a function of the parameters. Since the area is proportional to the volume sampled, the decrease of $\Delta H_{0}$ with $A$ is reasonable. On the other hand, the number density impacts the total number of SNe Ia in a similar manner as the area/volume. Given this dependence, we can perform a simple 2D polynomial fit on the parameters:
\begin{equation}\label{eq:fitting}
    \Delta H_{0} = a_{00} + a_{10}\frac{1}{A} + a_{01}\frac{1}{n_{sn}} +  a_{20}\frac{1}{A^{2}} + a_{11}\frac{1}{An_{sn}} + a_{02}\frac{1}{n_{sn}^{2}}
\end{equation}
using the Scikit-learn package \cite{scikit-learn} and the result is shown as the dotted line in the middle and right hand side panel of Fig.~\ref{fig:for_survey}. We can see that the fit is quite a good match to the calculations. In Table 3, we summarize the fitting parameters for a few values of $R$. We note that there is a caveat for the fitting result since, formally, the constant term should vanish when the survey area approaches infinity for any fixed density of SN Ia, but it does not in our fit. The problem is the data we use for the fit is based on noisy measurements, and that we only have data for a physical range of $A$ (less than the full sky). Given the current fitting accuracy and the fact that the survey area can never exceed that of the full sky, the current result can be used as a reasonable approximation for practical applications. For an ideal isotropic survey with isotropic coverage in the angular direction, the result can be used to provide a quick estimate for $\Delta H_{0}$. We note that the coefficients for the second order term are much smaller compared with the first order terms, indicating that a simple volume scaling may already be sufficient for an approximation.

\begin{table}
\centering
\begin{tabular}{lllllll}
\hline
R[$h^{-1}$Mpc]  & $a_{00}$ & $a_{10}$ & $a_{01}$ & $a_{20}$ & $a_{11}$ & $a_{02}$ \\
\hline
250 &  0.27 & 673 & 6.15e-8  & -5.72e5 & 4.87e-4  & -7.16e-14 \\
350 &  0.15 & 332 &  1.86e-8 & -2.74e5 & 2e-4 & -2.36e-14 \\
450 &  0.098 &  174 &  7.88e-9 & -1.33e5 &  8.44e-5  & -1.14e-14 \\
\hline
\end{tabular}
\caption{Fitting parameters for the dependence of $\Delta H_{0}$ on parameters $A$ and $n_{sn}$ using Eq. (\ref{eq:fitting}). Parameter $A$ is in unit of deg$^{2}$, while $n_{sn}$ is the number density of dark matter halos used in the analysis with a unit of [Mpc]$^{-3}$. }
\label{table}
\end{table}

\subsection{Implications for future SN Ia survey: realistic case}

\begin{figure}
\begin{center}
\includegraphics[width=7.5cm]{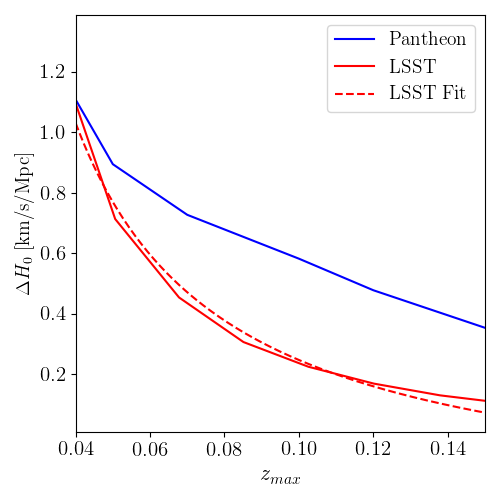}
\caption{$\Delta H_{0}$ forecast based on the LSST SN Ia observations using the UNIT simulation (red solid). We choose a LSST SQ case as detailed in \cite{Garcia_2020} and present the result as a function of $z_{max}$, the maximum of redshift in the $H_{0}$ measurement using the distance ladder. The current result from the Pantheon compilation is also shown (blue solid). The red dashed line shows a simple dependence on $1/z_{max}$. }
\label{fig:lsst}
\end{center}
\end{figure}

As another example, we can forecast $\Delta H_{0}$ from future SN observations such as LSST \cite{LSST_2009}, using the distance ladder method. Following \cite{Garcia_2020}, we predict the expectation assuming a 5-year LSST SQ survey, which can observe 110k SN events within an angular coverage of 18000 deg$^2$ and a maximum redshift of 0.35. Since higher redshift may confront redshift evolution and model dependency, the distance ladder usually truncates at e.g. $z<0.15$, we also restrict the analysis below this limit. In Figure \ref{fig:lsst}, we present the expected $\Delta H_{0}$ as a function of $z_{max}$. For comparison, we also display the current result from Pantheon compilation. The result shows that a 5 year complete survey may reduce the sample variance of $H_{0}$ to $\sim0.1$ km s$^{-1}$Mpc$^{-1}$, a quarter of the current estimate. In addition, we also fit the LSST result with a simple scaling relation $\Delta H_{0}=0.052/z_{max}-0.27$ as shown as the red line in the figure. Note that this fitting relation only uses measurements with redshift below 0.15 and should not extend to higher redshift regimes.

\subsection{Large scale modulation}

\begin{figure*}
\begin{center}
\includegraphics[width=8.5cm]{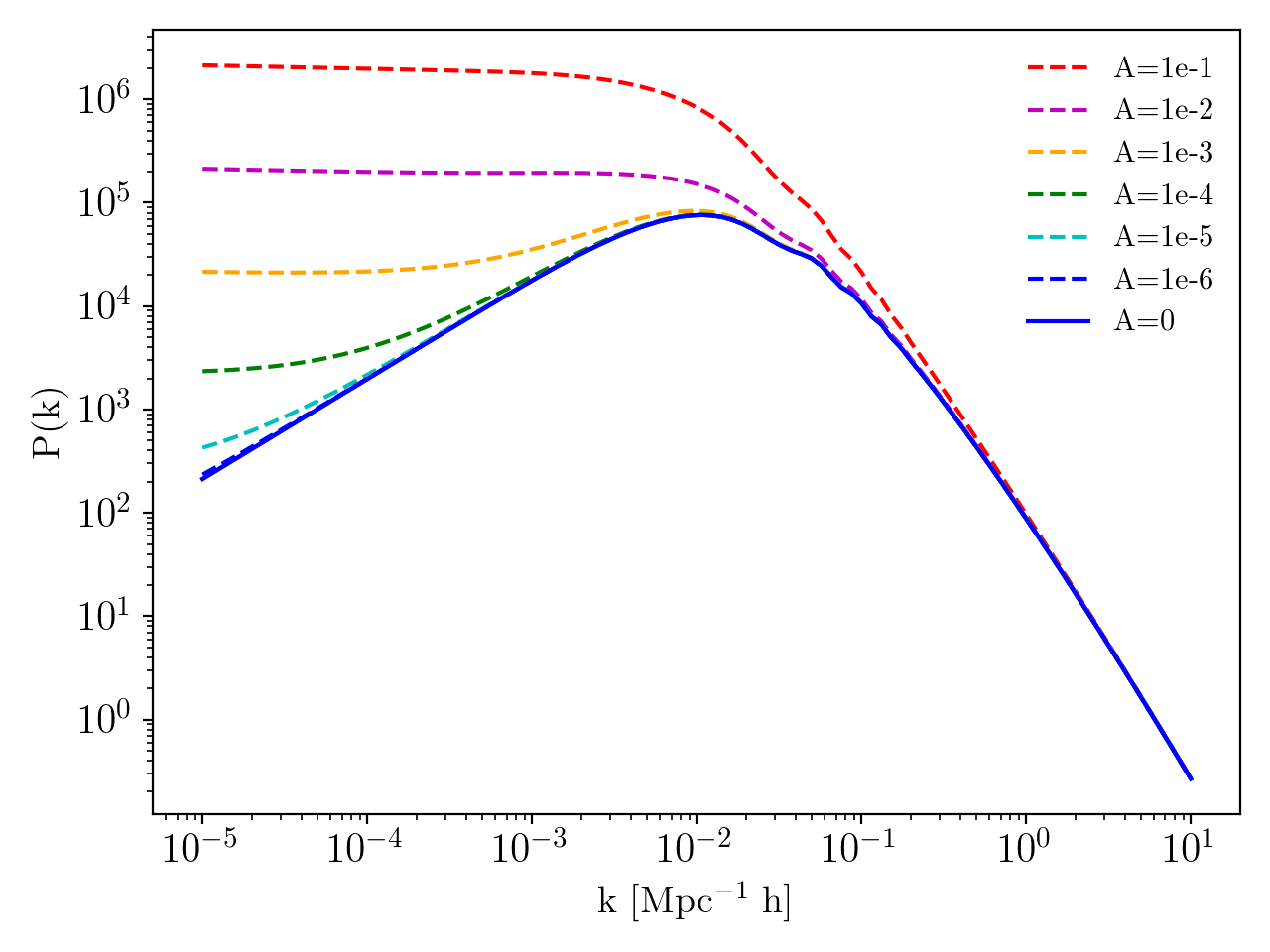}
\includegraphics[width=8.5cm]{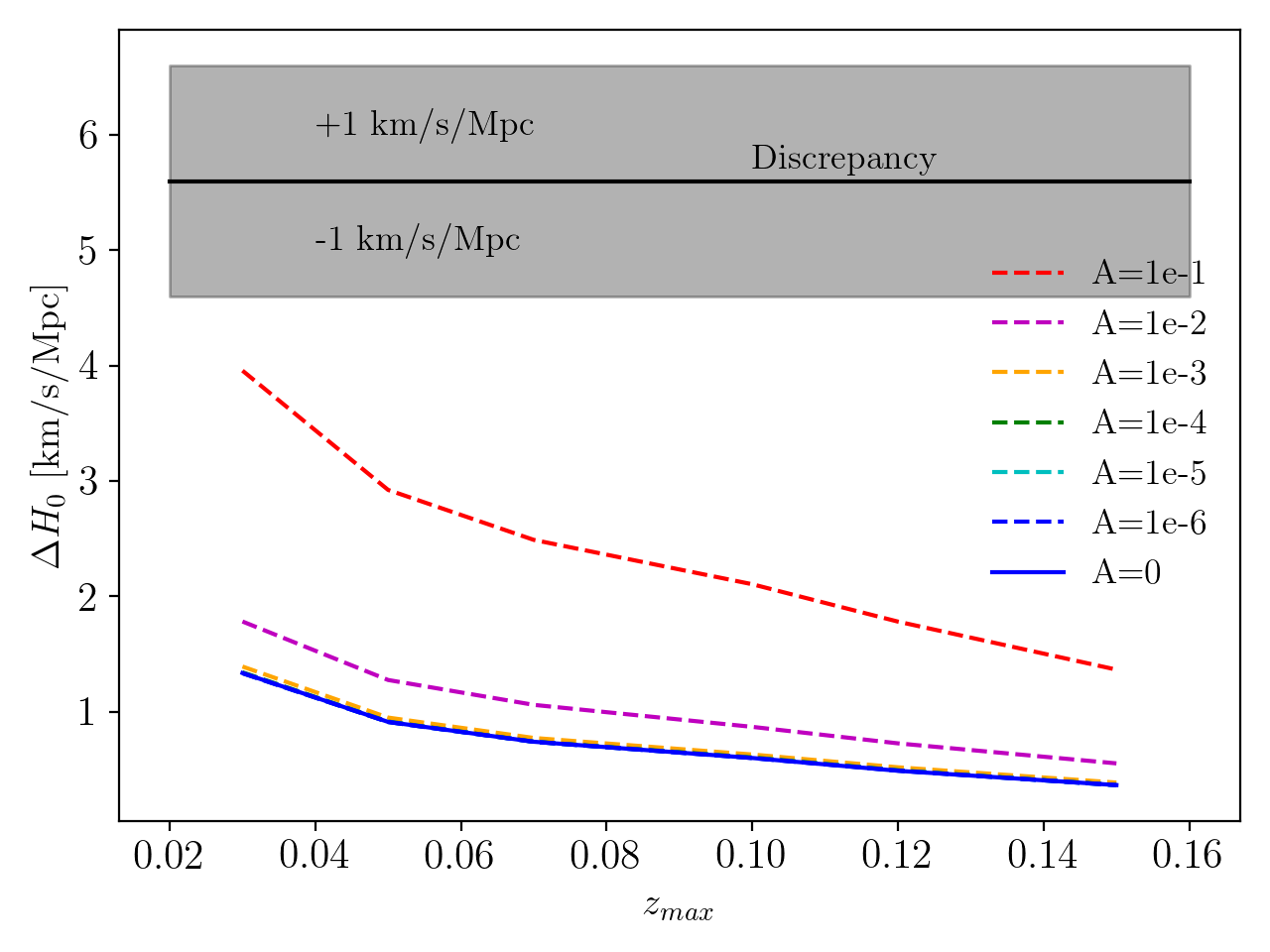}
\caption{\textbf{Left:} modulated matter power spectrum with different values of $A$. This toy model is designed to affect large scale more than small scale. \textbf{Right:} The resultant estimate of $\Delta H_{0}$ as a function of $z_{max}$ for different values of $A$. The horizontal line denotes the current $H_{0}$ discrepancy.}
\label{fig:pk_modulation}
\end{center}
\end{figure*}

In the velocity correlation function method, as in all of the methods to calculate the sample variance, the matter power spectrum is of critical importance. In particular, increasing the large-scale velocity power spectrum will increase the sample variance. Primordial non-Gaussianity of density perturbations affects the halo mass function and bias on large scale, leaving a $k^{-2}$ divergent signal in the large-scale power spectrum \cite{Dalal_2008, Slosar_2008, Matarrese_2008, Valageas_2010}. This does not, however, affect the velocity power spectrum as biased objects are expected to still trace the matter velocity field. Another examples of such effect is the curvaton model of the inflationary theory \cite{Erickek_2008a, Erickek_2008b}, which is proposed to explain the hemispherical power asymmetry from CMB observations \cite{Eriksen_2004}. This model introduces a large-amplitude superhorizon perturbation. It is possible that additional models may lead to excess power in the velocity power spectrum, increasing the sample variance.

In order to explore how much extra power is required to increase the sample variance and explain the current Hubble tension, we introduce a toy model to modulate the velocity power spectrum on large scales and investigate the implications on the $H_{0}$ sample variance. We multiply the matter power spectrum by a simple function $f = \frac{A}{k}+1$, where $A$ is a parameter that determines the amplitude of the modulation. The left panel of Fig.~\ref{fig:pk_modulation} shows the power spectrum when we change the value of parameter $A$. Then we use this modulated power spectrum and redo the calculation of $\Delta H_{0}$ using the velocity correlation function method as described in previous sections. The resultant $\Delta H_{0}$ is shown in the right panel of Fig.~\ref{fig:pk_modulation}. For comparison, the un-modulated result ($A=0$) is also shown. Increasing $A$ increases the variance of $H_{0}$, as expected given the calculation of velocity correlation function in Section \ref{sec:vcf}. The horizontal line shows the current $H_{0}$ offset of $\sim5.6$ km/s/Mpc between distance ladder and CMB measurements. For mild levels of modulation, $\Delta H_{0}$ does not increase significantly. For an extreme model with $A=0.1$ at $z_{max}=0.15$, the sample variance $\Delta H_{0}$ is up to $1.4$ km/s/Mpc, comparable and slightly larger than the total error budget using distance ladder. This uncertainty can reduce the $H_{0}$ tension below $4\sigma$. However, we should note that such an extreme model is not only modulating the velocity power spectrum at large scale, but changing the whole shape of the velocity power spectrum. Solving the full $H_{0}$ tension relying on sample variance and an excess of large-scale power does not look feasible given how much the velocity power spectrum would have to increase, and would produce a tension with other measurements, such as the CMB dipole or RSD for example \cite{Planck_CMB_dipole}. We can explicitly test this effect by looking at the velocity dispersion. Given our simple toy model, a large enough $\Delta H_{0}$ of a few km/s/Mpc to bring the tension within $1\sigma$ requires the parameter $A\sim O(1)$. Compared with the unmodulated power spectrum, this boosted model can increase the average amplitude of the peculiar velocity using Eq (\ref{eq:dispersion}) from $\sim$300 km/s to a few thousands km/s, i.e. a boost of roughly a factor of ten. In linear regime, we know that the velocity field is related to the density field via
\begin{equation}
v_{r}=-\beta \frac{\partial}{\partial r} \nabla_{\mathbf{r}}^{2}\delta,
\end{equation}
where $\nabla_{\mathbf{r}}^{2}$ is the inverse Laplacian and the subscript ``r" denotes the radial direction. If we assume that the underlying density field does not change, the boost of velocity field may purely and linearly come from the structure growth through parameter $\beta$. This leads to an increase of the linear growth rate by an order of magnitude, which significantly violates the current observations.

\section{Discussion and Summary}

We have extended our previous (paper I) comparison of methods to estimate the sample variance of $H_{0}$ by including an additional method based on the velocity correlation function. In this method, perturbation theory is used to construct a covariance matrix for the radial peculiar velocities of SNe Ia, and realisations are drawn from a mutivariate Gaussian using this matrix. We compare this with a method based on N-body simulations, and find that the replacement of the analytic peculiar velocity correlations with the non-linear simulation based correlations did not significantly change the estimate of $\Delta H_{0}$, indicating the robustness of the velocity correlation function using perturbation theory and the dominance of linear scales in the calculation. Given the extra $k^2$ dependence of the velocity power spectrum compared to that of the density, this is expected. This method allows the sample variance $\Delta H_{0}$ to be determined without the need for simulations or approximations about the geometry of the survey, such as that it is consistent with a spherical region.

Our analysis using numerical simulations also serves as an independent test for the importance of velocity bias. By selecting halos above some mass scale and compare the results using dark matter particles, we find consistent results showing that velocity bias is very close to unity and not important for such analyses. In addition, we test contributions from non-linear scale by simply applying a non-linear correction to the matter power spectrum while keeping the linear relation between density field and velocity field. This is not a full description of non-linear velocity field, but does show that the non-linear correction is not significant for the estimate of $\Delta H_{0}$.

One of the key findings of this paper is to provide a method to determine an effective volume for a set of SNe Ia data, and we apply this to determine the effective volume of the Pantheon SN 1a sample. It is obvious that the volume must be smaller than the maximum redshift of SNe Ia in the sample, since the SN observation only probes the space-time along the LoS and the density inhomogeneity is not representative of the whole spherical volume defined by the redshift. By comparing methods, we can quantitatively estimate this effective volume by comparing the error contribution to $H_{0}$ from the current Pantheon data with a spherical volume of different radii. Our result shows that the current distance ladder measurement is equivalent to a spherical volume with a radius of about 220 $h^{-1}$Mpc, roughly half the distance to $z=0.15$. When we investigate the locally perturbed background within a cosmological model such as the Lemaitre-Tolman-Bondi (LTB) model, one needs to be careful about the volume or scales that the SNe Ia data can actually sample. In terms of the $H_{0}$ uncertainty itself, our estimate of the sample variance is consistent with the LTB or void-based analysis that they are not able to resolve the tension, see e.g. \cite{LTB-H0, Huterer_2023}

Given the estimate of $\Delta H_{0}$ made using the velocity correlation function method, we can anticipate how the uncertainty of $H_{0}$ is affected by the volume covered, total number, or distribution of the SNe Ia. Using the simulated halo catalog, we study the dependence of $\Delta H_{0}$ on the volume and number density of SNe Ia. In this case, the sample variance is purely from peculiar velocity and we find a clear dependence on the volume and number density. We consider a possible future SNe Ia survey with isotropic distribution and fit the dependence on volume and number density with a second order polynomial model allowing fast calculations. We find that $\Delta H_{0}$ is more sensitive to the total area than number density: as SNe Ia surveys become sample variance limited covering a wide angular regions will become important. With a half-sky survey, or LSST-like survey in the future, we can expect that the sample variance of $H_{0}$ will decrease to 0.1 km s$^{-1}$Mpc$^{-1}$, a significant improvement compared with the latest measurement.

Within the framework of the velocity correlation function method, we can additionally investigate the impact of boosting the velocity power spectrum at large scales. We apply a simple toy model to modulate the power spectrum at large scales. As expected, the boost of large scale power can increase the final estimate of the $H_{0}$ sample variance. However, to mitigate or solve the tension, it requires an extreme model which may beyond the parameter space allowed by the current observational data. This analysis further demonstrates the severity of the current $H_{0}$ tension.

\begin{acknowledgments}

We thank the anonymous reviewers for their careful reading of our manuscript and their many insightful comments and suggestions that have significantly improved this paper. Research at Perimeter Institute is supported in part by the Government of Canada through the Department of Innovation, Science and Economic Development Canada and by the Province of Ontario through the Ministry of Colleges and Universities. ZZ is supported by NSFC (12373003), the National Key\&D Program of China (2023YFA1605600), and acknowledges the generous sponsorship from Yangyang Development Fund. ZD is supported by NSFC (12273020) and National Key R\&D Program of China (2023YFA1607800, 2023YFA1607802).

This research was enabled in part by support provided by Compute Ontario (computeontario.ca) and the Digital Research Alliance of Canada (alliancecan.ca).

\end{acknowledgments}

\appendix

\bibliography{SN-SSC}

\end{document}